\documentclass{PoS}

\title{Polyakov Loop Correlations at Large N}

\ShortTitle{Polyakov Loop Correlations at Large N}

\author{\speaker{Herbert Neuberger}%
         \thanks{Supported by the NSF, award PHY-1415525.}\\
        Rutgers University, Department of Physics and Astronomy, Piscataway, NJ 08855\\
        E-mail: \email{neuberg@physics.rutgers.edu}}

\abstract{I describe a study of the 
two-point single-eigenvalue distribution 
correlation function of Polyakov loops in the 
confined phase of four dimensional SU(N) YM theory at large N.
The reasons for the interest in this correlation function
are explained. Analytical and numerical
results are presented. Brief conclusions are drawn. 
}

\FullConference{The 32nd International Symposium on Lattice Field Theory\\
                 23-28 June, 2014\\
                 Columbia University New York, NY}

\begin{document}

\section{Introduction}

Quantum smeared loops~\cite{smear} in pure $SU(N)$
YM theory incur a qualitative change when they are
dilated from a small to a large size. Controlling
the crossover is a basic problem. At infinite $N$, 
the crossover collapses to a point, becoming
a large-$N$ phase transition. Below the transition
asymptotically free perturbation
theory holds and above it a description by an effective string theory (EST) is valid. 
One would like to match these two ranges at the
transition point and test that the matching works
using lattice gauge theory. The problem one faces
is that EST in its simplest form 
requires the loop to be smooth and contractible 
loops on the
lattice have kinks. Non-contractible loops (Polyakov
loops) do not have kinks but their expectation value
is zero due to a $Z(N)$ symmetry. So, the question
is whether one can find a large $N$ phase transition
in the correlation function of two Polyakov loops at
large $N$. Since this object is subleading in the large $N$ expansion, the issue is subtle~\cite{main}.

\section{Large $N$ transition in contractible 
Wilson loops.}

The single-eigenvalue distribution of smeared Wilson loops 
undergoes a ``compactification'' transition on the unit circle at $N=\infty$ \cite{phasetransition}. 
Below is an example of a $6\times 6$ 
smeared Wilson loop of size
$0.6$ Fermi at $N=29$.
\begin{center}
\includegraphics[width=0.5\textwidth]{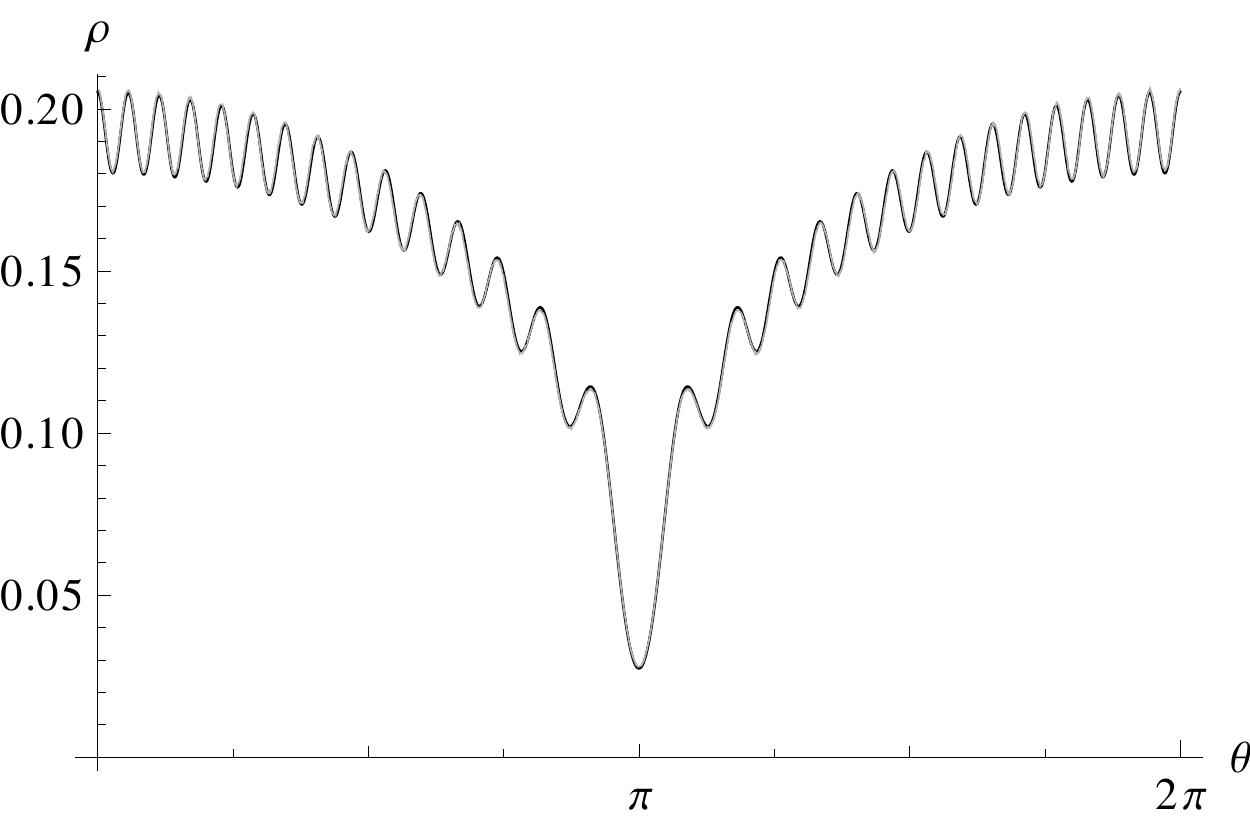}
\end{center}
I would like to 
calculate approximately $\sigma$ in units of $\Lambda_{QCD}$ for $N=\infty$ 
by matching EST (effective string theory) and PT (YM perturbation theory) at the transition point. This is a natural matching point: at $N=\infty$ the parallel transporter round the loop does not
reach the vicinity of the -1 element of the $SU(N)$ group with probability
one for smaller loops, while, for larger loops, the support of the eigenvalues
of this parallel transporter is the entire 
group manifold. This indicates that the 
perturbative asymptotic expansion in the logarithm of loop size is a valid approximation for loops smaller than the transition size, but not for
larger loops, where the compactness of 
the group manifold is detected and full exponentiation
of the Lie algebra is necessary. EST is expected to be an asymptotic description
for large loops with validity possibly extending to the entire regime of loop
sizes exceeding the critical size.

Previous work~\cite{wilsonloops} has led me to the conclusion that 
long distance behaviour is described by EST, but
EST works in too limited a way for loops with 
kinks. EST requires smooth loops and one needs
a situation where it works best. I also have 
to maintain the ability to test the matching 
procedure against Monte Carlo data. So, I need 
to work with smooth loops on the lattice. 
Polyakov loops are the single available option. 
Hence, I look for an $N=\infty$ transition 
associated with Polyakov loops within the low temperature phase.

\section{Setup}
I first define my notation. 
The Polyakov parallel transporter is denoted by
\begin{equation}
U_P (x)={\cal P} e^{i\oint_{x_4}^{x_4} A_4 ({\vec x},\tau) d\tau}. 
\end{equation}
Here ${\vec x}$ denotes the space component of the four-vector $x$. 
In the continuum limit, the quantum smeared 
$U_P(x)$ is a matrix
with operator valued entries which satisfies the
same unitarity conditions a unitary c-number matrix would.
The set of its eigenvalues $e^{i\theta_k}$ is gauge invariant.  
The character of the parallel transporter in the irreducible
representation $R$ is given by  
$P_R ({\vec x})= \frac{1}{d_R} \chi_R (U_P(x))$. It is
independent of $x_4$. 
The two point correlation function of two Polyakov loops at
two space points depends only on their spatial separation  $r$ and is denoted by  $G_R(r)=\langle P_R(0) P_{\overline R} (r)\rangle$.

The two point function is positive (the theta
parameter in the YM action is set to zero) and its logarithm is a useful quantity:
$W_R(l,r)=\log G_R(r)$, where $l$ is the length of the
compact direction and $r$ the loop separation. 
As an example of possibly the strongest EST prediction consider the quantity ${\cal F}_R(l)$:
\begin{equation} 
{\cal F}_R(l)=\lim_{r\to\infty} \partial^2 W_R(l,r)/\partial l \partial r
\end{equation}
For $1\le n \le N-1$, the ``$N$-ality", we consider 
${\cal F}_R(l) = \sigma_n {\hat F}_R(l\sqrt{\sigma_n})$.
This is the case where EST makes its strongest prediction in our context: 
\begin{equation}{\hat F}_R(x)=1+c_1/x^2+c_2/x^4+c_3/x^6+...,
\end{equation}
 where the $c_{1,2,3}$ are three universal, calculable numbers, independent of $R$ and $n$~\cite{aharoni}. Taking the large size limits in different ways typically produces weaker results. 
 
\section{2D YM model}
In the context of non-analyticities 
generated by taking $N$ to infinity 
in the 't Hooft prescription, 
previous work has shown that two dimensional
YM theory provides a representative of the ``universality class'' associated with the large $N$ transition. Therefore, I first study the 
eigenvalue-eigenvalue correlation for 
Polyakov loop matrices in 2D YM. Specifically, 
I compute a two point function of eigenvalue densities $\rho^{(1)}(\theta; U)=\frac{2\pi}{N} \sum_{k=1}^N \delta_{2\pi} (\theta-\theta_k)$.

One starts from the ``propagator'' ~\cite{gross}
\begin{equation}
Z_N(U_{P_1},U_{P_2}|t)=\sum_R \chi_R (U_{P_1}) e^{-\frac{t}{2 N} C_2(R)} 
\chi_{\overline R} (U_{P_2}),
\end{equation}
intending to calculate
\begin{equation}
~\langle\rho^{(1)}_1(\alpha)\rho^{(1)}_2(\beta)\rangle_c=\int 
dU_{p_1} dU_{p_2} \\ \rho^{(1)}_1(\alpha)\rho^{(1)}_2(\beta) 
[Z_N(U_{p_1},U_{p_2}| t) -1]
\end{equation}
This can be done using the character expansion~\cite{twod} 
in terms of hook-type Young diagrams ${(p,q)}$
\begin{equation}
 \rho^{(1)}(\theta;U)=1+\frac{1}{2N} \lim_{\epsilon\to 0^+} \sum_{p=0}^{N-1}\sum_{q=0}^\infty
(-1)^p e^{-\epsilon(p+q+1)} [ e^{i(p+q+1)\theta} \chi_{(p,q)} (U) +
e^{-i(p+q+1)\theta} \chi_{\overline{(p,q)}} (U)].
\end{equation}
For simplicity, I will restrict myself to odd $N$.
Using $C(p,q)=(p+q+1)(N-\frac{p+q+1}{N} +q-p)$, I obtained
\begin{equation}\langle \rho^{(1)}_1(\alpha)\rho^{(1)}_2(\beta)\rangle_c =\frac{1}{N^2} 
\sum_{p=0}^{N-1} \sum_{q=0}^\infty
(-1)^p e^{-\frac{t}{2N} C(p,q) }\cos[(p+q+1)(\alpha-\beta)]
\end{equation}
Taking the large $N$ limit gives:
\begin{equation}N^2\langle \rho^{(1)}_1(\alpha)\rho^{(1)}_2(\beta)\rangle\sim
\Re \sqrt{\frac{N}{t}} ue^{-\frac{t}{2}+\frac{t}{2N^2}} \int\frac{dx}{\sqrt{2\pi}} 
e^{-\frac{N}{2t} x^2 +\frac{1}{2t} x^2} \frac{1+u^N e^{-N(x+\frac{t}{2})+\frac{t}{2}}}{1+ue^{-x-\frac{t}{2}+
\frac{t}{2N}}}\frac{1}{1-ue^{-\frac{t}{2}}}, 
\end{equation}
where $u=\exp[i(\alpha-\beta)]$.

The answer consists of the sum of a rapidly oscillating piece and a non-oscillating piece
\begin{equation}\frac{1}{2}\frac{\sinh\frac{t}{2} 
\cos\phi }{\sinh^2\frac{t}{2}+\sin^2\phi},
\end{equation}
where
$\phi=\alpha-\beta$. The expression differs from
the universal form for random hermitian matrix models~\cite{eynard}, likely because of
the absence of the potential term of the latter. 
There is no large $N$ transition 
separating regimes of small $t$ and 
large $t$.
The approximate large $N$
formula is compared with the exact finite $N$ formula below with the solid line showing the exact result. One sees that the approximate large $N$ expression 
deteriorates when $N$ decreases, when $\phi\approx k\pi, k\in Z$ and when $t$ is small relative to 1.
\vskip 0.25cm
$N=11$, $t=0.3,1,5$ from top to bottom: 
\begin{center}
\includegraphics[width=0.4\textwidth]{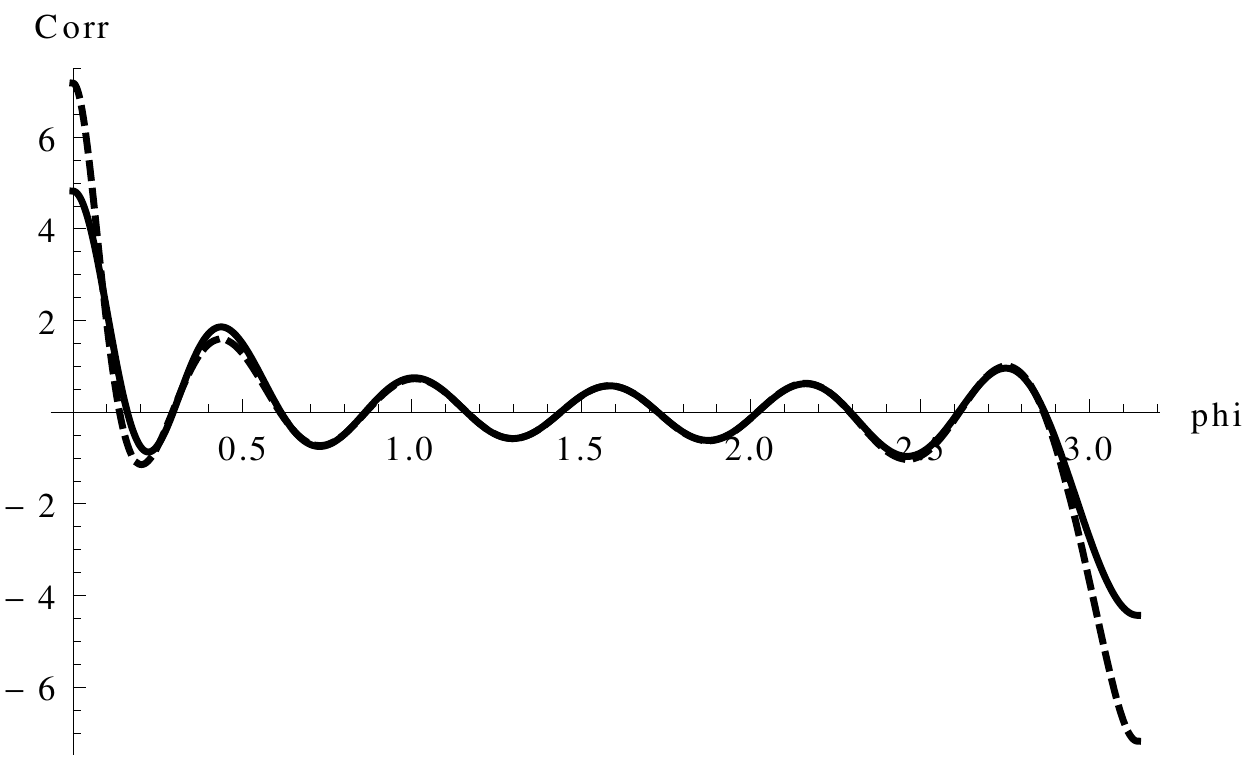}
\end{center}
\begin{center}
\includegraphics[width=0.4\textwidth]{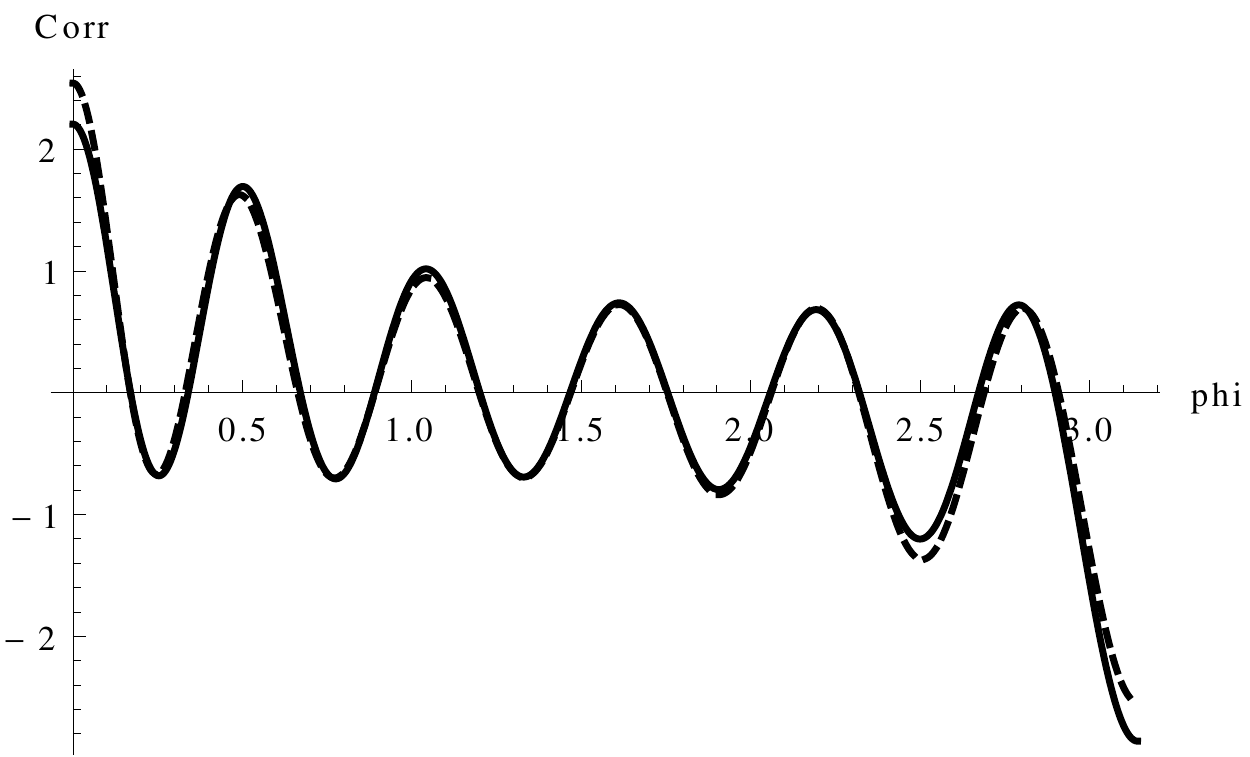}
\end{center}
\begin{center}
\includegraphics[width=0.4\textwidth]{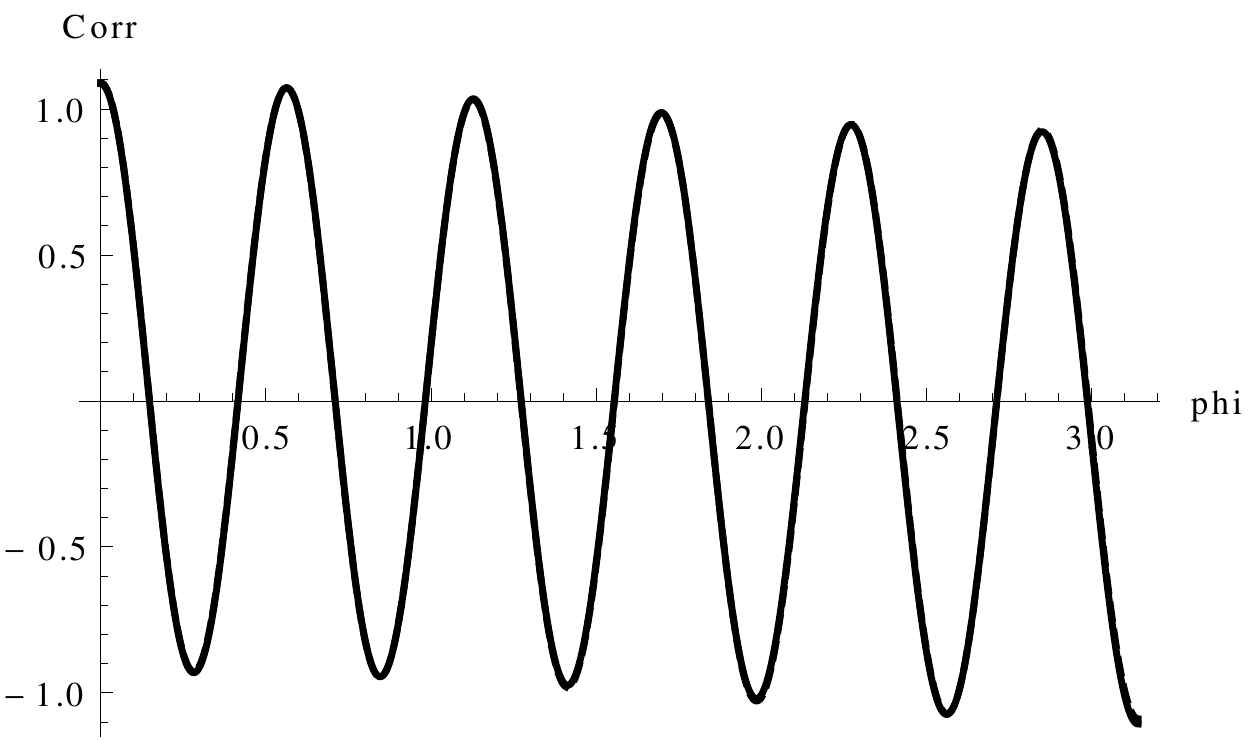}
\end{center}
\vskip 0.25cm
$N=29$, $t=0.3,1,5$ top to bottom: 

\begin{center}
\includegraphics[width=0.4\textwidth]{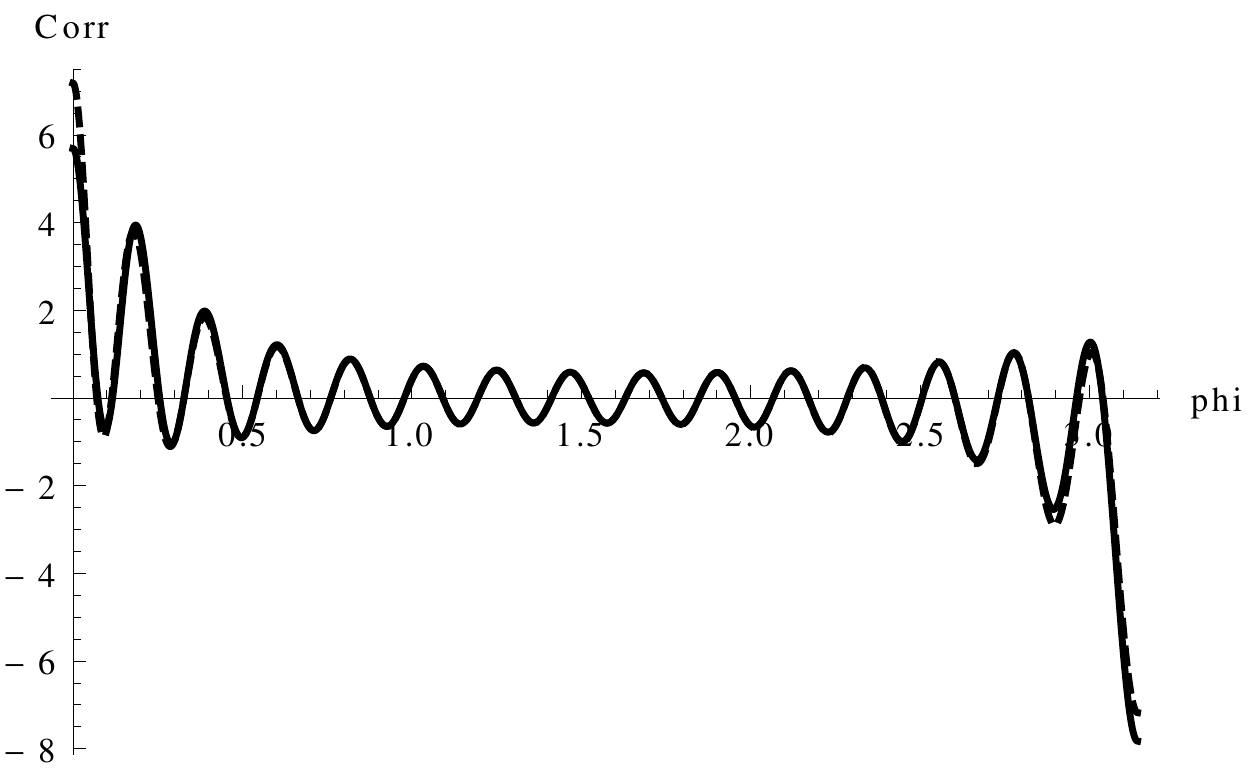}
\end{center}
\begin{center}
\includegraphics[width=0.4\textwidth]{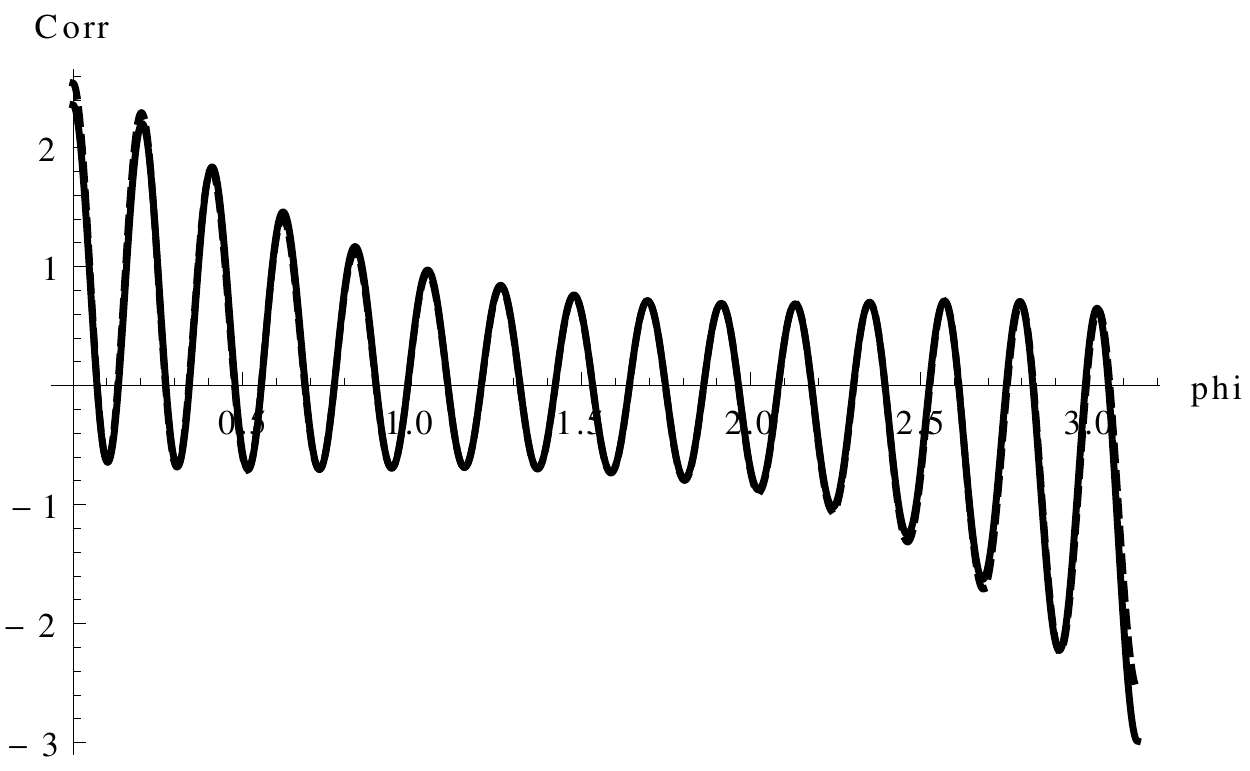}
\end{center}
\begin{center}
\includegraphics[width=0.4\textwidth]{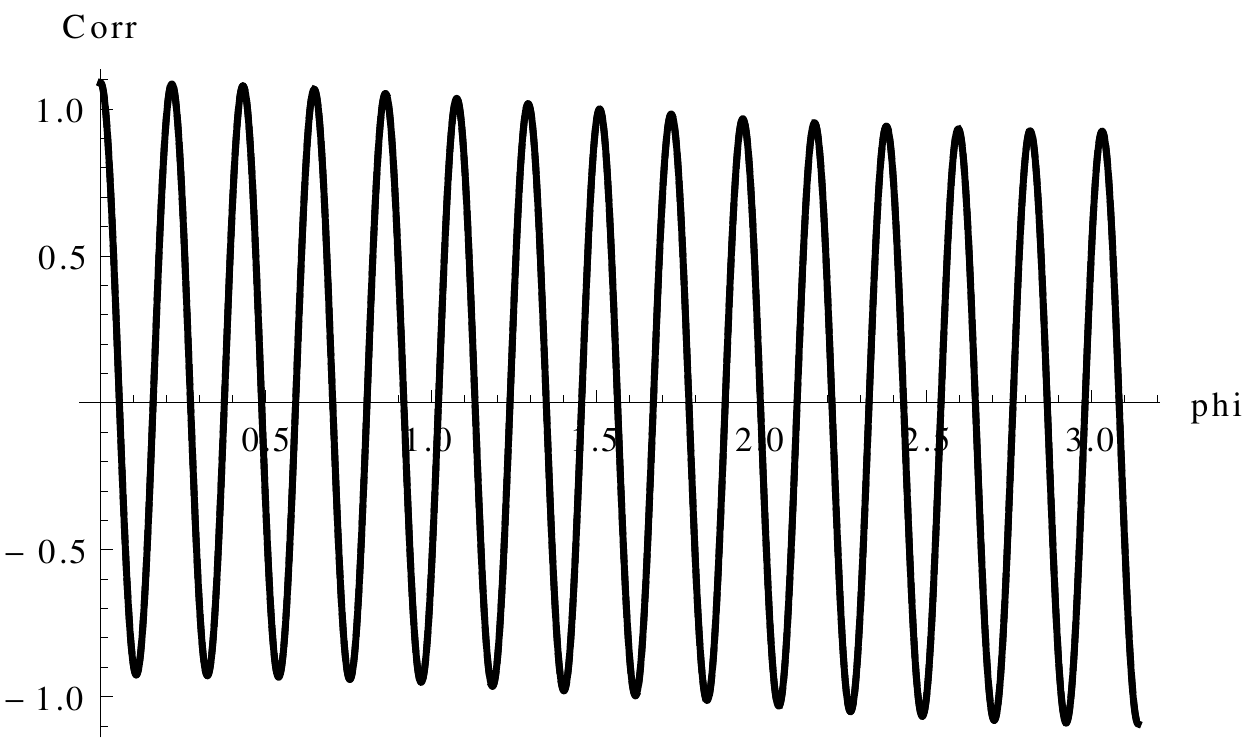}
\end{center}
\vskip 0.25cm
We see that the large $N$ expression I derived analytically checks against finite $N$ expressions
evaluated numerically. 

\section{4D results}
For finite $N$, there is no reason for $\rho^{(2)}$ to depend only on the
angle difference since the $Z(N)$ symmetry only provides invariance under simultaneous
shifts of $\alpha$ and $\beta$ by $2\pi k/N$. 
Initial simulations were 
done collecting two dimensional histograms in the $\alpha,\beta$ plane.
Is was found that within practical numerical accuracy collapsing the
histograms along constant $\alpha-\beta$ lines did not loose any information.
This means that we may as well redefine 
$\rho^{(2)}$:
\begin{equation}
\rho^{(2)}(\alpha-\beta)=
\frac{N}{2\pi}\int_{-\pi/N}^{\pi/N} d\theta 
\langle \rho^{(1)}_1(\alpha+\theta)\rho^{(1)}_2(\beta+\theta)
\rangle_c 
\label{rho2}
\end{equation}
producing a 
$\rho^{(2)}$ depending only on the angle difference on account of the $Z(N)$ symmetry.

An example of the outcome of a Monte Carlo simulation 
in 4D is shown below. 
In addition to raw data, I show a smoothed curve obtained by a cubic spline
smoothing method. The method of smoothing consists of a minimization
of a weighted combination of an average of the curve curvature and deviation from the data. 
The smoothing procedure is quite ad-hoc, and only serves to produce curves to guide the eye.
\vskip .5cm 
\begin{center}
\includegraphics[width=0.4\textwidth]{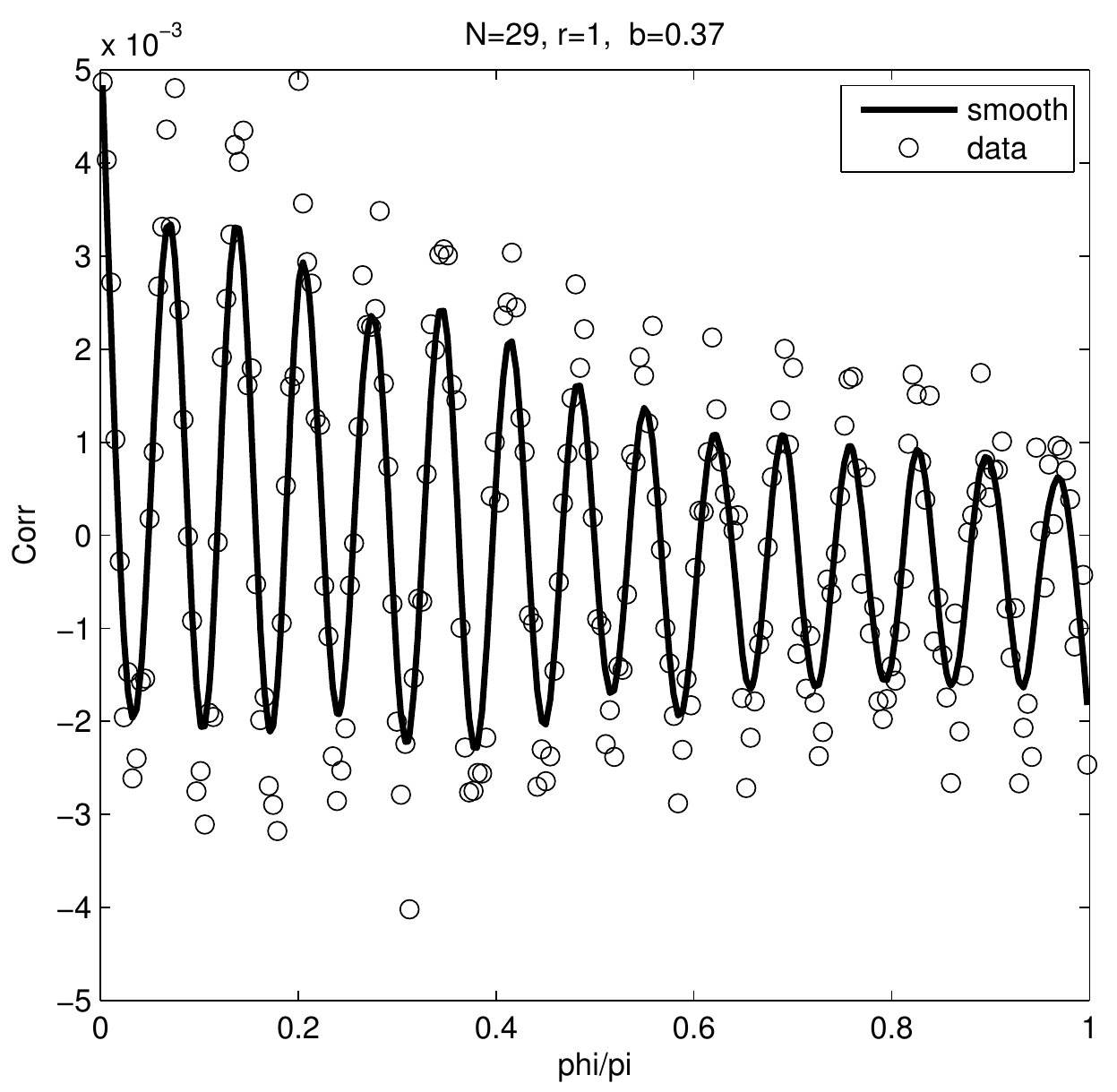}
\end{center}
\begin{center}
\includegraphics[width=0.4\textwidth]{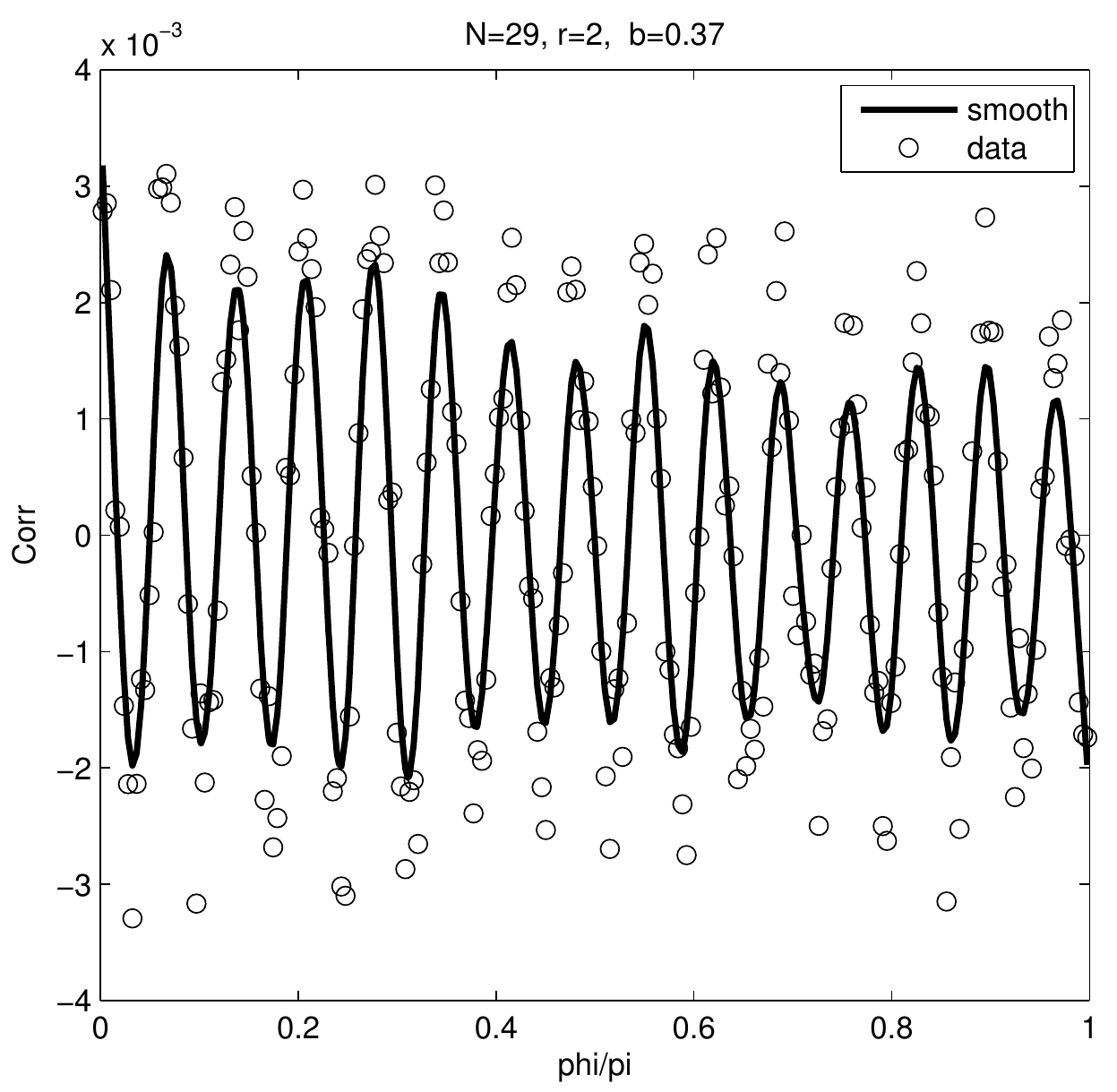}
\end{center}
\begin{center}
\includegraphics[width=0.4\textwidth]{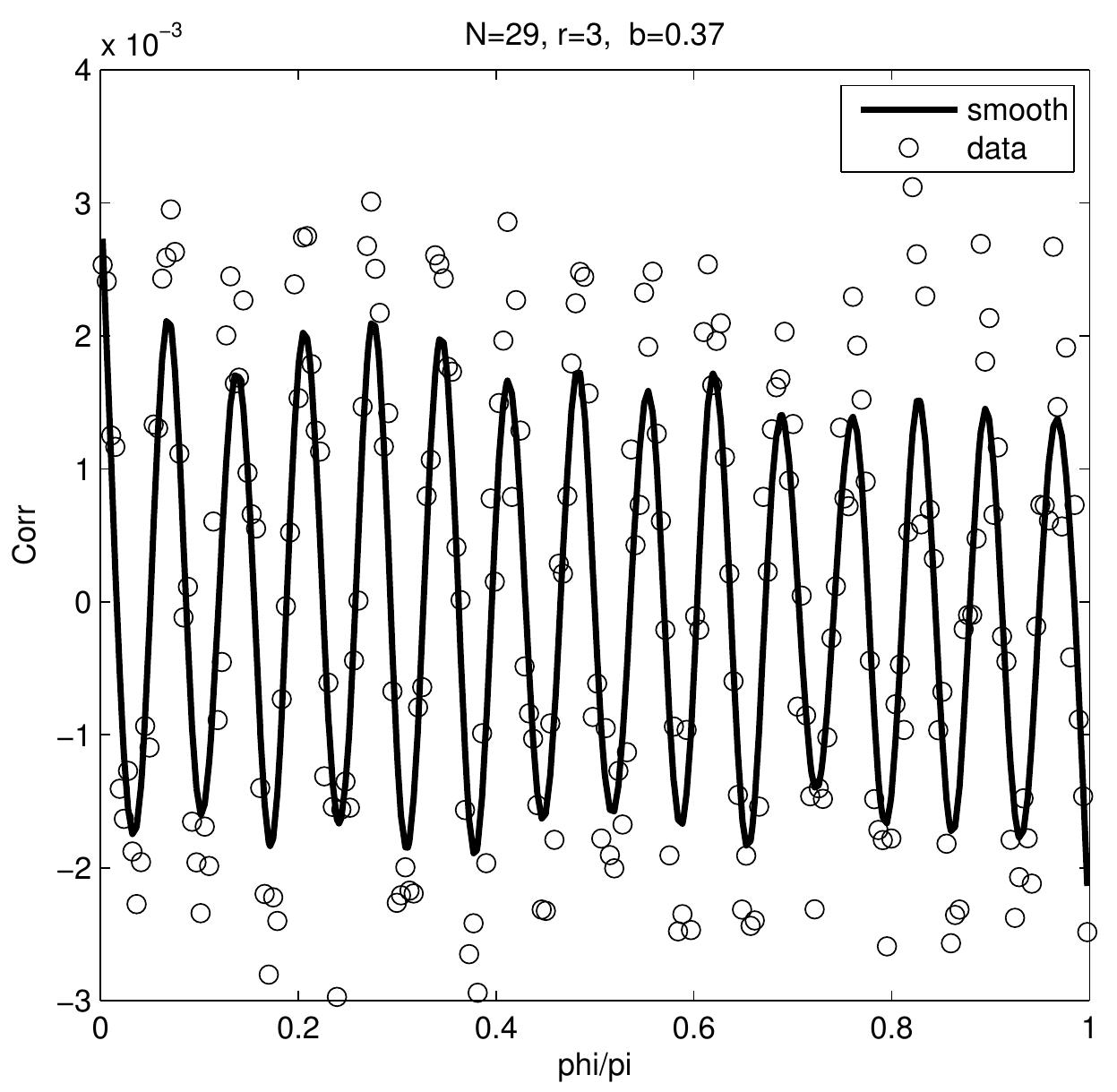}
\end{center}
\vskip .5cm
These results were obtained for $N=29$ and rescaled
't Hooft coupling $b(\equiv\frac{\beta}{2 N^2})=0.370$ at
separation $r=1,2,3$ in lattice units from top to bottom.
Only half of the angular range is shown.  

Qualitatively, the curves resemble their two dimensional counterparts, but the noise is large. The
results indicate no
large $N$ phase transition in this observable in 4D. 
I have not ruled out that the redefinition in eq. (\ref{rho2}) hid a transition. It would be numerically expensive to do this.

\section{Conclusions and Outlook}

There is no large $N$ phase transition for large enough Polyakov loops
as their separation is varied. 
To get a large $N$ transition one would have to shrink the compact direction, 
while maintaining the system in the confined phase. 
This phase would be metastable. 
This may be possible using quenching techniques and would be of theoretical
interest also in another respect~\cite{polch}. 

Other observables, involving the analogue of the
2D YM ``vertex'', and which combine different windings
might be of interest and could potentially provide
better candidates for observables undergoing large $N$
phase transitions. For more details I refer to~\cite{main}.

\end{document}